\documentclass[a4paper]{article}
\usepackage{Odyssey2020}
\usepackage{epsfig,amssymb,amsmath}
\usepackage{xcolor}
\usepackage{float}
\usepackage{multirow}
\usepackage{booktabs}
\usepackage{dcolumn}
\usepackage[colorlinks=true,linkcolor=black,anchorcolor=black,citecolor=black,filecolor=black,menucolor=black,runcolor=black,urlcolor=black]{hyperref}
\ninept

\usepackage{tikz}
\usetikzlibrary{bayesnet}
\renewcommand{\edge}[3][]{ %
  \foreach \x in {#2} { %
    \foreach \y in {#3} { %
      \path (\x) edge [->,#1] (\y) ;%
    } ;
  } ;
}

\usetikzlibrary{arrows,shapes,backgrounds,positioning,fit}

\def\Vmat{\mathbf{V}}

\def\ND{\mathcal{N}}

\DeclareMathOperator*{\argmax}{argmax}

\def\R{\mathbb{R}}

\def\Lset{\mathcal{L}}

\def\Lmat{\mathbf{L}}
\def\Bmat{\mathbf{B}}
\def\Wmat{\mathbf{W}}

\def\Smat{\mathbf{S}}

\def\Imat{\mathbf{I}}

\def\Xmat{\mathbf{X}}

\def\yvec{\mathbf{y}}

\def\xvec{\mathbf{x}}

\def\const{\text{const}}

\def\inblue#1{{\color{blue}#1}}

\def\inred#1{{\color{red}#1}}

\newcommand{\inv}[1]{%
\ifx{#1}{\Imat}           
  #1
\else
  {#1}^{-1}
\fi
}

\tikzstyle{cbox} = [rectangle,draw=blue!100,thick,align=center,rounded corners = 3pt]
\tikzstyle{lbox} = [rectangle,draw=blue!100,thick,align=left,rounded corners = 3pt]
\tikzstyle{ccircle} = [circle,draw=blue!100,thick,align=center,inner sep = 0]
\tikzstyle{ctext} = [rectangle,align=center,inner sep = 4pt]
\tikzstyle{ltext} = [rectangle,align=left,inner sep = 4pt]
\tikzstyle{solder} = [circle,draw,fill,inner sep = 0, minimum size = 3pt]

\title{Probabilistic embeddings for speaker diarization}

\makeatletter
\def\name#1{\gdef\@name{#1\\}}
\makeatother
\name{{\em Anna Silnova$^1$, Niko Br\"ummer$^2$, Johan Rohdin$^{1,2}$, Themos Stafylakis$^2$, Luk\'a\v{s} Burget$^1$}
			}

\address{$^1$Brno University of Technology, FIT, IT4I CoE, Czechia\\
$^2$Omilia - Conversational Intelligence, Athens, Greece \\
{\small \tt{ \{isilnova,rohdin,burget\}@fit.vutbr.cz}, \{nbrummer,tstafylakis\}@omilia.com}}

\begin{document}
\maketitle

\begin{abstract}
Speaker embeddings (x-vectors) extracted from very short segments of speech have recently been shown to give competitive performance in speaker diarization. We generalize this recipe by extracting from each speech segment, in parallel with the x-vector, also a diagonal precision matrix, thus providing a path for the propagation of information about the quality of the speech segment into a PLDA scoring backend. These precisions quantify the uncertainty about what the values of the embeddings might have been if they had been extracted from high quality speech segments. The proposed \emph{probabilistic embeddings} (x-vectors with precisions) are interfaced with the PLDA model by treating the x-vectors as hidden variables and marginalizing them out. We apply the proposed probabilistic embeddings as input to an agglomerative hierarchical clustering (AHC) algorithm to do diarization in the DIHARD'19 evaluation set. We compute the full PLDA likelihood `by the book' for each clustering hypothesis that is considered by AHC. We do joint discriminative training of the PLDA parameters and of the probabilistic x-vector extractor. We demonstrate accuracy gains relative to a baseline AHC algorithm, applied to traditional x-vectors (without uncertainty), and which uses averaging of binary log-likelihood-ratios, rather than by-the-book scoring.           
\end{abstract}

\section{Introduction}
We are interested in doing speaker diarization by extracting embeddings from very short (1.5s or shorter) speech segments, followed by clustering of the embeddings w.r.t.\ speaker. To process a speech recording for diarization, we use the following cascade: (i) a deep neural net (DNN) to extract embeddings from short consecutive segments of speech; (ii) a probabilistic discriminant analysis (PLDA) backend that processes the embeddings to compute likelihoods for hypothesized speaker clusters; (iii) agglomerative hierarchical clustering (AHC) to greedily find an approximately optimal clustering. This is one of the standard ways of doing diarization, but we propose modifications to all three stages.    

Intuitively, for speaker diarization purposes, the quality of speech segments could depend on factors like speech duration, reverberation, microphone placement, signal-to-noise ratio and contamination by overlapped speech. We do not propose here to explicitly design extractors for such quality factors, but we rely instead on the ability of DNNs to learn to extract relevant information from the input. 

By treating the embeddings as hidden variables, we derive a mechanism to propagate the quality information, that is supplied by the embedding extractor, into the scoring and clustering backend. This gives an augmented embedding extractor that computes from each speech segment not only an x-vector, but also a precision matrix that quantifies the uncertainty about what the value of the x-vector might have been if it had been extracted from a high quality speech segment. We show how to modify both PLDA and AHC to make use of this extra information.

We discriminatively train the new embedding extractor jointly with the PLDA parameters, using a novel multiclass cross-entropy criterion formed by the average log-posterior probabilities for the correct clustering of $n$-tuples of embeddings. In this paper we use $n=8$.  

The setup for this work is derived from our winning submission to
DIHARD'19 challenge, where x-vectors were clustered by Bayesian HMM (BHMM)~\cite{landini2019but}. Although in this paper we use the 
simpler AHC clustering, our proposed mechanisms of quality propagation could be applied to BHMM in future.

Section~\ref{sec:priorwork} reviews prior work. Section~\ref{sec:theory} develops the theory to marginalize over the hidden embeddings and to score the modified PLDA backend. This sets the stage for discriminative training and AHC clustering in sections~\ref{sec:training} and~\ref{sec:ahc}. Finally, experiments and results in section~\ref{sec:experiments} demonstrate the theory.\footnote{\href{https://github.com/bsxfan/probabilistic_embeddings}{https://github.com/bsxfan/probabilistic\_embeddings} 
}

\section{Prior work}
\label{sec:priorwork}
Space constraint precludes a review of the extensive literature on speaker diarization. The focus and novelty of this paper is the probabilistic embedding mechanism, which we place into context relative to similar prior work. 

In i-vector speaker recognition, the quality effects of duration and phonetic variability were propagated  into the PLDA backend in~\cite{CRIM_ivector_uncertainty,Sandro_uncertainty}. We cannot build on these works, since they relied on the i-vector posterior uncertainty as defined by the \emph{generative} i-vector extraction model, while the x-vector extractor that we use here is \emph{discriminative}. Our method allows propagation of more general quality factors.          

In our previous work~\cite{Brummer2018}, we propagated scalar-valued uncertainties extracted from i-vector/x-vector magnitudes into a heavy-tailed PLDA backend, while the embedding extractor remained as-is. Our new work is more general in the following aspects: (i) The embedding extractor is modified and trained to extract (ii) vector-valued uncertainty; (iii) The hidden variable, around which the propagation hinges, is now the embedding, rather than the speaker identity variable as in~\cite{Brummer2018}.

In~\cite{hib} \emph{posterior probability  distributions} for embeddings given images are proposed. Similar to our approach, they treat the embeddings as hidden variables. In contrast, we work with \emph{likelihood distributions}, the advantages of which are discussed later.       

In~\cite{McCree_pseudoLH} an embedding extractor was discriminatively trained for diarization. They did not use uncertainty propagation and their training criterion (pseudolikelihood) was different from ours. It should be noted that both their pseudolikelihood~\cite{Dawid_discrete_PSR} and our cross-entropy are proper scoring rules~\cite{PSR} that encourage good calibration of the trained models.

\section{Theory}
\label{sec:theory}
In this section we present a novel way to derive a discriminatively trainable embedding extractor from a generative model. The generative approach facilitates a principled interface between the extracted embeddings and the PLDA backend. Although the whole model is generative, we show how to effectively ignore parts of the model so that it can be discriminatively trained. After training, the same model can be used at runtime for diarization. We make our modelling assumptions subject to tractability and computational efficiency constraints. 

 Let $\Smat=\{s_t\}_{t=1}^n$ denote an $n$-tuple of speech segments, for which we consider the following generative model. First, a speaker clustering hypothesis~\cite{SPP}, denoted $\Lmat$, is generated from some prior distribution, $P(\Lmat)$. Then $\Lmat$ generates an $n$-tuple of \emph{hidden embeddings}, $\Xmat=\{\xvec_t\}_{t=1}^n$, where $\xvec_t\in\R^D$. Finally each speech segment, $s_t$, is generated from the associated $\xvec_t$: 
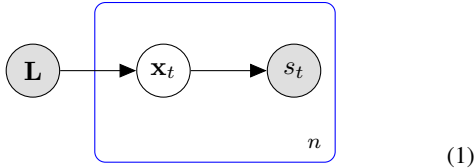
\begin{align}
\label{dg:model1}
\begin{tikzpicture}
\node[obs] (C) {$\Lmat$};
\node[latent, right = of C] (x) {$\xvec_t$};
\node[const, outer sep = 20pt] at(x-|x) (xdummy) {$$};
\node[obs, right = of x] (s) {$s_t$};
\plate[draw=blue,inner sep = 5pt] {} {(xdummy)(s)} {$n$};
\edge {C}{x};
\edge {x}{s};
\end{tikzpicture}
\end{align} 
The $s_t$ are always observed, $\Lmat$ is observed at training time, but not at runtime, while the $\xvec_t$ are always hidden. Notice that according to this model, the $\xvec_t$ would be the \emph{ideal} features for inference of $\Lmat$ from $\Smat$, because of the conditional independence:
\begin{align}
P(\Lmat\mid\Xmat,\Smat) &=  P(\Lmat\mid\Xmat)
\end{align} 
Unfortunately, these ideal features cannot be directly observed or computed and we have to resort to marginalization to compute the \emph{clustering  posterior}: 
\begin{align}
\label{eq:Cpost}
P(\Lmat\mid\Smat) &= \int  P(\Lmat,\Xmat\mid\Smat) \,d\Xmat
\end{align}
The rest of this section is devoted to constructing a model, with associated tractable algorithms, that enables computation of~\eqref{eq:Cpost}.

\subsection{The partition prior}
\label{sec:prior}
The Bell number, $B_n$, gives the number of ways that $n$ items can be partitioned (clustered). For octets, we have $B_8=4140$. The support of the partition prior, $P(\Lmat)$, can include up to $B_n$ discrete possibilities, although we might sometimes consider priors with fewer possibilities---e.g.\ if we know there are fewer than $n$ speakers. 

In this paper, we use a Chinese restaurant process (CRP) to define $P(\Lmat)$. The CRP assumes that the sequence of speaker labels are exchangeable and that time plays no role. More details of our CRP prior are given later. In future work, to model dependency between consecutive segments, $P(\Lmat)$ could be defined by a Markov model~\cite{landini2019but}, or perhaps by a distance-dependent CRP~\cite{DDCRP}. 

\subsection{The PLDA model}
We introduce the PLDA backend via a more detailed view of model~\eqref{dg:model1}, which now includes also $n$ \emph{speaker identity variables}, $\yvec_i\in\R^d$. For simplicity in defining this model, we let the maximum number of speakers be equal to the number of speech segments, $n$. The clustering hypothesis is represented here as set of \emph{speaker labels}, $\Lmat=\{\ell_t\}_{t=1}^n$, where each $\ell_t\in\{1,\ldots,n\}$.  
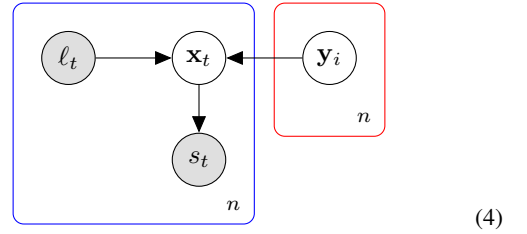
\begin{align}
\label{dg:model2}
\begin{tikzpicture}
\node[latent] (x) {$\xvec_t$};
\node[obs, left = of x] (label) {$\ell_t$};
\node[latent, right = of x] (y) {$\yvec_i$};
\node[const, outer sep = 15pt] at(x-|x) (xdummy) {$$};
\node[const, outer sep = 15pt] at(y-|y) (ydummy) {$$};
\node[const, outer sep = 15pt] at(label-|label) (ldummy) {$$};
\node[obs, below = 2 em of x] (s) {$s_t$};
\plate[draw=blue,inner sep = 5pt] {} {(xdummy)(s)(ldummy)} {$n$};
\plate[draw=red,inner sep = 5pt] {} {(ydummy)} {$n$};
\edge {label}{x};
\edge {x}{s};
\edge {y}{x};
\end{tikzpicture}
\end{align} 
Only a subset of the speakers might produce observations: There is an arc from every $\yvec_i$ to every $\xvec_t$, but when $\ell_t=k$, then only $\yvec_k$ generates observation $\xvec_t$.      

The label set, $\Lmat=\{\ell_t\}_{t=1}^n$, requires some further explanation. In the case of octets for example, there are $B_8=4140$ possible ways to cluster $8$ elements. But the label set, if it were unconstrained, would have $8^8$ (almost 17 million) possibilities. We can however arrange a deterministic, one-to-one relationship between $\Lmat$ and the possible clusterings by constraining $\Lmat$ to be a \emph{restricted growth string}~\cite{orlov}, subject to the constraints:
\begin{align}
\label{eq:lgs}
\ell_1&=1 &&\text{and} & \ell_{t+1} &\le 1 + \max(\ell_1,\ldots,\ell_t)
\end{align}  
The $\yvec_i$ are generated IID from the standard $d$-dimensional multivariate Gaussian, while the $\xvec_t$ are generated from the $D$-dimensional conditional Gaussian:  
\begin{align}
\label{eq:samplex}
\xvec_t\mid(\yvec_k,\ell_t=k)\sim \ND(\Vmat\yvec_k,\Wmat^{-1})
\end{align} 
where $\Vmat$ is the $D$-by-$d$ speaker factor loading matrix and $\Wmat$ is the $D$-by-$D$ within-speaker precision matrix. If $D>d$, this model is usually referred to as simple PLDA (SPLDA), while if $D=d$ and $\Vmat$ is of full rank, the model is referred to as the \emph{two covariance model}~\cite{SPP}, where $\Vmat\Vmat'$ is the between-speaker covariance.  

\subsection{Marginalization}
The marginalization of each $\xvec_t$ can be done independently of its siblings, by computing the \emph{speaker identity likelihood}, given segment $s_t$:
\begin{align}
\label{eq:sil}
\begin{split}
P(s_t\mid\yvec) &= \int P(s_t,\xvec_t\mid\yvec) \,d\xvec_t \\
&= \int P(\xvec_t\mid \yvec)P(s_t\mid \xvec_t) \,d\xvec_t
\end{split}
\end{align} 
where $\yvec$ is the identity variable of the speaker of segment $t$. The first factor in the final integrand is Gaussian, given by~\eqref{eq:samplex}. If we can force the second factor, $P(s_t\mid \xvec_t)$, the \emph{likelihood} for $\xvec_t$ given $s_t$, to be Gaussian as a function of $\xvec_t$, then the integral can be computed in closed form. We shall do exactly that by imposing a constraint on our model. But first, we need to consider (carefully) the following factorization of the likelihood, which can be done without loss of generality:
\begin{align}
\begin{split}
\label{eq:gsump}
P(s_t\mid \xvec_t) &=  \inred{\frac{P(s_t)}{g(s_t)}}\times\inblue{g(s_t)\frac{P(\xvec_t\mid s_t)}{P(\xvec_t)}} \\
&=\inred{h(s_t)}\;\times\inblue{f(s_t,\xvec_t)}
\end{split}
\end{align}  
where we have introduced the real-valued functions $f,g,h$, such that $g,h$ are strictly positive and $f$ is non-negative. The function $f(s_t,\xvec_t)$ is a formal representation of our embedding extractor, a trainable DNN, that implements the mapping: $s_t\mapsto f(s_t,\cdot)$, where the output (a likelihood distribution) is the \emph{probabilistic embedding} that is extracted from $s_t$. Both $P(s_t\mid \xvec_t)$ and $f(s_t,\xvec_t)$ are valid forms for the likelihood for $\xvec_t$ given $s_t$. The arbitrary functions $g,h$ serve to express in full generality the unnormalized nature of likelihood distributions. 

The difference between the probabilistic embeddings of~\cite{hib} and ours is shown by:
\begin{align}
f(s_t,\xvec_t) &= g(s_t)\frac{P(\xvec_t\mid s_t)}{P(\xvec_t)}
\end{align}
We use the likelihood, $f(s_t,\xvec_t)$, while~\cite{hib} uses the posterior, $P(\xvec_t\mid s_t)$. The likelihood is  unencumbered by the necessarily complex embedding prior, $P(\xvec_t)$, which in our case is defined by the PLDA backend. If your embeddings are any good at separating classes, then $P(\xvec_t)$ needs to reflect this and will take the form of a mixture with well-separated components. This is also true of the posterior, $P(\xvec_t\mid s_t)$, and indeed in~\cite{hib} they are forced to work with complex probabilistic embedddings in mixture form.\footnote{In the i-vector--PLDA cascade, the simple, single Gaussian i-vector prior is \emph{inconsistent} with a view of the whole cascade as a single generative model. In such a model, the i-vector prior would be defined by the PLDA backend, which is impractically complex.} In the likelihood form, the complex prior is effectively `divided out' and we can work with simpler embedding distributions. We achieve more freedom in the form of the embedding by this decoupling from the backend. 

As will be shown below, $h(s_t)$  cancels in our scoring formulas and therefore plays no role at runtime or during discriminative training, affording us the luxury of tractable calculations involving only $f$. The function, $h$ carries all the complexities of the distribution of the raw input speech and by side-stepping it, we enjoy the discriminative privilege of not having to model the input. Nevertheless, discriminative training of a parametric form of $f(s_t,\xvec_t)$ (as we do) has some implications for~\eqref{eq:gsump} which are discussed further in appendix~\ref{sec:appendix}.

We can now introduce our modelling constraint. Since $\xvec_t$ is hidden, we have considerable freedom in choosing its nature (discreet or continuous), its dimensionality and its relationship to $s_t$. We let $f(s,\xvec)$ be Gaussian in $\xvec$, with mean and precision matrix that are arbitrarily complex functions of $s$.  We represent the Gaussian, $f(s_t,\xvec_t)$, as:
\begin{align}
\label{eq:fdef}
f(s_t,\xvec) &= \exp\left[-\frac12\xvec'\Bmat_t\xvec+\xvec'\Bmat_t\hat\xvec_t\right]
\end{align}
where $\hat\xvec_t\in\R^D$ and $\Bmat_t$ (positive semi-definite precision matrix) are functions of $s_t$. In summary, our proposed, trainable DNN embedding extractor does: $s_t\mapsto\hat\xvec_t,\Bmat_t$. This is very similar to the encoder of a variational autoencoder~\cite{VAE}. Note however that the Gaussian~\eqref{eq:fdef} is unnormalized and in general not even normalizable, unless $\Bmat_t$ is positive definite (invertible). Our recipe works in general for non-invertible $\Bmat_t$, as long as the PLDA parameter, $\Wmat$, is invertible.  

After some algebra to solve the Gaussian integral, plugging~\eqref{eq:samplex}, \eqref{eq:gsump} and~\eqref{eq:fdef} into~\eqref{eq:sil} gives:
\begin{align}
\label{eq:marg}
\begin{split}
P(s_t\mid\yvec)\propto
\exp&\left[-\frac12\yvec'\Vmat'(\Wmat-\Wmat(\Wmat+\Bmat_t)^{-1}\Wmat)\Vmat\yvec\right.\\
&\;\;\;\;+ \left. \yvec'\Vmat'\Wmat(\Wmat+\Bmat_t)^{-1}\Bmat_t\hat\xvec_t\right]
\end{split}
\end{align}   
where we have omitted factors that are independent of $\yvec$, including $h(s_t)$.

\subsection{Diagonalization}
\label{sec:diag}
One way to avoid the $\mathcal{O}(D^3)$ computational requirements of~\eqref{eq:marg}, is to specialize to the two-covariance model, with $D=d$ and full rank, square $\Vmat$. Then, without loss of generality, we can linearly transform the hidden $\xvec_t$, so that $\Vmat\Vmat'$ and $\Wmat$ are mutually diagonalized. First, transform $\xvec_t$ so that $\Vmat\Vmat'=\Imat$. This is followed by pure rotation that diagonalizes $\Wmat$, while preserving $\Vmat\Vmat'=\Imat$. We are now free to let $\Vmat=\Imat$. Finally, we also constrain the extracted embedding precision, $\Bmat_t$ to be diagonal. We can now express~\eqref{eq:marg} as the product of univariate Gaussians:    
\begin{align}
\label{eq:dsil}
P(s_t\mid\yvec) &\propto \prod_{j=1}^D \exp\left[\frac{w_j b_{jt}}{w_j +b_{jt}}(\hat x_{jt}y_j-\frac12y_j^2)\right]
\end{align}   
where $w_j>0$ and $b_{jt}\ge0$ are diagonal elements of $\Wmat$ and $\Bmat_t$, while $y_j$ and $\hat x_{jt}$ are components of $\yvec$ and $\hat\xvec_t$. If $b_{jt}>0$, then Gaussian $j$ has mean $\hat x_{jt}$ and variance $\frac{w_j +b_{jt}}{w_j b_{jt}}$. High quality components with $b_{jt}\gg w_j$, will have low uncertainty, with variance saturated at $1/w_j$. Conversely, low quality components with $b_{jt}\ll w_j$ can have arbitrarily large uncertainty. At the limit, $b_{jt}=0$, the associated component $\hat x_{jt}$ will be completely ignored in any downstream processing, as will be shown below.

\subsection{The clustering posterior}
\label{sec:cpost}
The clustering (partition) posterior~\eqref{eq:Cpost} can be expressed as:
\begin{align}
\label{eq:CPost}
P(\Lmat\mid\Smat) &= \frac{P(\Lmat)\prod_{i=1}^nP(\Smat^{(i)}\mid\Lmat)}{\sum_{\tilde\Lmat}P(\tilde\Lmat)\prod_{i=1}^nP(\Smat^{(i)}\mid\tilde\Lmat)}
\end{align}
where the summation in the denominator is over all $B_n$ possibilities allowed by~\eqref{eq:lgs} and where $P(\Smat^{(i)}\mid\Lmat)$ is the joint distribution for all speech segments\footnote{If no segments are attributed to speaker $i$, we conveniently let $P(\Smat^{(i)}\mid\Lmat)=1$.} attributed to speaker $i$ by label set $\Lmat$. We expand:
\begin{align}
\label{eq:PSi}
P(\Smat^{(i)}\mid\Lmat) &= \int P(\yvec) \prod_{t\in\Lmat_i} P(s_t\mid \yvec)\, d\yvec
\end{align}
where $\Lmat_i$ is the set of speech segment indices attributed to speaker $i$ by $\Lmat$. For $P(s_t\mid \yvec)$, we can plug the RHS of~\eqref{eq:marg} or~\eqref{eq:dsil} into~\eqref{eq:PSi}, because the omitted factors cancel in~\eqref{eq:CPost}. The cancelled factors include $\prod_{t=1}^n h(s_t)$, which occurs in the numerator and in each term of the denominator. Recalling that $P(\yvec)$ is standard Gaussian, the integrand in~\eqref{eq:PSi} is a product of Gaussians so that the integral can be solved in closed form. For the diagonal case, we find after some algebra:
\begin{align*}
\log P(\Smat^{(i)}\mid\Lmat) &= \frac12\sum_{j=1}^D \left(\frac{\bar a_j^2}{1+\bar b_j} - \log(1+\bar b_j)\right) + \const
\end{align*}
where
\begin{align}
\label{eq:bars}
\bar a_j &= \sum_{t\in\Lmat_i} \frac{w_j b_{jt}}{w_j + b_{jt}}\hat x_{jt}, &
\bar b_j &= \sum_{t\in\Lmat_i} \frac{w_j b_{jt}}{w_j + b_{jt}}
\end{align}  
The constant is independent of $\Lmat$ and cancels in all our scoring and training recipes. Notice again that if $b_{jt}=0$, then $\hat x_{jt}$ is effectively removed from the computation. Conversely, if $b_{jt}\gg w_j$, then the weight for $\hat x_{jt}$ saturates at $w_j$. If all $b_{jt}$ are sufficiently large, the result is the same as when traditional (fixed) embeddings, $\hat\xvec_t$, are just plugged into PLDA. The normalized, weighted sum~\eqref{eq:bars} may be interpreted as an attention mechanism~\cite{AisAYN}, where the $w_j$ are the queries, the $b_{jt}$ are the keys and the $\hat x_{jt}$ are the values.

\subsubsection{The posterior computation recipe}
\label{sec:recipe}
In summary, to compute the clustering posterior for $n$-tuples, we need: 
\begin{itemize}
	\item A DNN that serves as probabilistic embedding extractor that does $s_t\mapsto\{\hat x_{jt},b_{jt}\}_{j=1}^D$. We assume that the final stage of the extractor for $\hat\xvec_t$ is a trainable linear transform, so that we may assume the extracted embeddings will have been transformed to be compatible with a diagonalized two-covariance backend.
	\item A diagonalized, two-covariance PLDA model that supplies the within-speaker precision parameters, $\{w_j\}_{j=1}^D$. 
	\item A precomputed clustering prior, $P(\Lmat)$, that respects~\eqref{eq:lgs}. The prior can be stored in a table of size $B_n$, containing log probabilities.
	\item A precomputed sparse $n$-by-$(2^n-1)$ matrix, with $0/1$ entries, that can be used to efficiently do the summations to compute $\bar a_j$ and $\bar b_j$ in~\eqref{eq:bars}, for all possible subsets $\Lmat_i$. From these, we can compute $\log P(\Smat^{(i)}\mid\Lmat)$ for every possible subset.
	\item Another precomputed sparse $B_n$-by-$(2^n-1)$ matrix, with $0/1$ entries, that can be used to efficiently accumulate $\sum_{i=1}^n \log P(\Smat^{(i)}\mid\Lmat)$, for every possible value of $\Lmat$. After addition of $\log P(\Lmat)$, a softmax of size $B_n$ computes the final posterior~\eqref{eq:CPost}. 
\end{itemize}

\section{Discriminative training}
\label{sec:training}
\def\Dset{\mathcal{D}}
We jointly train the embedding extractor and the PLDA parameters by minimizing the following $B_n$-way multiclass cross-entropy criterion, computed using~\eqref{eq:CPost}:
\begin{align}
\mathcal{C} &= - \sum_{(\Lmat,\Smat)\in\Dset} \log P(\Lmat\mid\Smat) 
\end{align} 
where $\Dset$ is a collection of supervised \emph{trials}, each containing an $n$-tuple of speech segments, $\Smat$, and the associated true speaker clustering, $\Lmat$. In this paper, we choose $n=8$ and we refer to the $n$-tuples as \emph{octets}. The selection of segments to compose octet trials is described in the section on experiments below. Note that $B_n=4140$, which gives a relatively large, but still tractable number of clustering hypotheses for each octet trial.  

To compute the clustering posterior~\eqref{eq:CPost}, we require also a clustering prior, $P(\Lmat)$. In this paper we use a Chinese restaurant process, which will be detailed later.


\section{Diarization with Agglomerative Hierarchical Clustering}
\label{sec:ahc}
At training time we can control $n$, the number of speech segments to cluster, so that $B_n$ remains tractable. At runtime, where $n$ may be a few hundred, $B_n$ far exceeds the number of atoms in the known universe. This makes an exact search for an optimal clustering hopelessly intractable. 

Agglomerative hierarchical clustering (AHC) is a greedy algorithm for finding approximately optimal clustering solutions. For speaker diarization, AHC is initialized with one cluster per segment. Each segment is represented by an x-vector (baseline), or probabilstic x-vector (proposed). At each iteration, one pair of clusters is joined, according to some local optimality condition computed using the x-vectors. When a stopping criterion is met, each of the final clusters is attributed to a different speaker.

The baseline AHC algorithm described below cannot be interpreted as a search for the maximum-likelihood (ML) clustering, because of the way that inter-cluster comparisons are computed (via binary log-likelihood-ratio averaging). In contrast, our proposed AHC algorithm \emph{can} be considered to be a greedy search for the ML solution. 

\subsection{Baseline AHC}
\label{sec:baseAHC}
The official Kaldi diarization recipe~\cite{sell2018diarization} implements the following clustering algorithm.
First, some PLDA model is used to calculate log-likelihood ratio verification scores as a similarity metric for each pair of x-vectors from the given test recording. The resulting pair-wise similarity matrix is the only input to the {\em unweighted average linkage} AHC (also known as UPGMA). At each stage of the algorithm the highest score is selected from the similarity matrix and two clusters corresponding to that score are merged. The row and column of the similarity matrix corresponding to a new cluster is computed as an average between the original scores of the two cluster components. 

The similarity score threshold $\sigma$ for stopping the AHC process is estimated for each recording separately using an {\em unsupervised linear calibration}~\cite{NikoDani_unsupcal}: a GMM with two univariate Gaussian components with shared variance is trained on all the scores from the similarity matrix. The two Gaussian components are assumed to be the score distributions corresponding to the same-speaker and different-speaker x-vector pairs. Therefore, $\sigma$ is set as the score for which the posterior probability of both components is 0.5 (i.e. decision threshold for the same/different-speaker maximum-a-posteriori classifier).

\subsection{By-the-book AHC}
\label{sec:bookAHC}
Our proposed AHC is a greedy maximization of the clustering log-likelihood: 
\begin{align}
\Lset(\Lmat) &= \log P(\Smat\mid\Lmat) = \sum_{i\in\text{clusters of $\Lmat$}} \log P(\Smat^{(i)}\mid \Lmat) 
\end{align}
Here $\Smat$ represents all the speech segments in the recording and $\Lmat$ is a clustering hypothesis for all these segments. Each term of $\Lset(\Lmat)$ can be computed by~\eqref{eq:bars}, up to a constant that is irrelevant when comparing hypotheses. $\Lmat$ is initialized to have each segment in a separate cluster. At each iteration, we do:
\begin{align}
\label{eq:localmax}
\Lmat \gets \argmax_{\Lmat'\xleftarrow{\text{join 2 clusters}}\Lmat} \Lset(\Lmat')
\end{align}
where $\Lmat'$ is restricted to merging a single pair of clusters in $\Lmat$. The iteration is stopped when the new hypothesis fails to give an improvement in $\Lset(\Lmat)$ that exceeds a preset threshold, $\sigma$. For greedy maximum likelihood, the correct threshold is $\sigma=0$, but we make it adjustable to help compensate for the fact that maximum likelihood ignores the prior and for any other mismatches between the model and the real data.   

We implement~\eqref{eq:localmax} using the log-likelihood-ratios, 
\begin{align}
\label{eq:Delta}
\Delta &= \Lset(\Lmat')-\Lset(\Lmat)    
\end{align}
where log-likelihood terms not in the to-be-merged clusters conveniently cancel. The pair with the highest $\Delta$ is joined in each iteration, except when the best one fails to exceed $\sigma$. For to-be-joined clusters $i$ and $j$, $\Delta$ is:
\begin{align*}
\log\frac{P(\Smat^{(i)},\Smat^{(j)}\mid \ell'_{1}=\dots=\ell'_{N^i+N^j})}
{P(\Smat^{(i)}\mid \ell_{i,1}=\dots=\ell_{i,N^i})P(\Smat^{(j)}\mid \ell_{j,1}=\dots=\ell_{i,N^j})}
\end{align*}
The $\ell_{i,k}\in\Lmat$ are the identical speaker labels for the $N^i$ segments of cluster $i$ and the same applies to cluster $j$. The $\ell'_k\in\Lmat'$ are the new labels for the joined cluster. $\Smat^{(i)},\Smat^{(j)}$ are the sets of speech segments assigned to each cluster. Likelihoods in the numerator and denominator are computed using (\ref{eq:bars}). The computation can be done reasonably fast, by  precomputing $\frac{w_j b_{jt}}{w_j + b_{jt}}\hat x_{jt}$  and $\frac{w_j b_{jt}}{w_j + b_{jt}}$. These quantities can be stored in matrices, where each row initially corresponds to a segment and later to a cluster. When two clusters are merged, the corresponding rows of the matrices are summed, so that each matrix will have one row fewer.

\section{Experiments and results}
\label{sec:experiments}
\subsection{Training and evaluation data}
The deep neural network (DNN)  x-vector extractor was trained on VoxCeleb 1 and 2 \cite{chung2018voxceleb2} with 1.2 million speech segments from 7146 speakers plus additional 5 million segments obtained with data augmentation.
Also, VoxCeleb data are used to train the baseline PLDA model.

Data from the AMI corpus \cite{CarlettaAMI} were used for training PLDA and probabilistic embedding extractor. We have split the data into training and cross-validation parts. Training part uses 75\% of the available data. The speakers between training and cross-validation sets are not overlapping.

Finally, we perform the diarization experiments on DIHARD 2019 development and evaluation data \cite{DIHARDEvalPlan,DIHARDCorpora}. Development set, in our case, was used for hyper-parameter tuning (such as selection of the AHC stopping threshold), otherwise, no development data were used for training system parameters. Performance is evaluated in terms of \emph{Diarization Error Rate} (DER). 

\subsection{Signal processing}\label{dereverb-WPE}
We used the weighted prediction error (WPE) \cite{nakatani-wpe-2010,Drude2018NaraWPE} method to remove late reverberation from the evaluation data. We estimated a dereverberation filter on short-time Fourier transform (STFT) spectrum for every 100 seconds block of an utterance. To compute the STFT, we used 32ms windows with 8ms shift. We set the filter length and prediction delay to 30 and 3 respectively for 16kHz. The number of iterations was set to 3.

\subsection{X-vector extractor} \label{xvecs}
We used the x-vector extractor from BUT's submission to the 2019 DiHard challenge \cite{landini2019but,LandiniICASSP20,DiezICASSP20}. The extractor was trained with the Kaldi toolkit~\cite{povey2011kaldi} using the SRE'16 recipe~\cite{snyder_kaldi_recipe} with the following modifications:
\begin{itemize}
    \item We used 40-dimensional filterbank features generated using 16kHz sampling frequency.
    \item The networks was trained for 6 epochs (instead of 3).
    \item We used 200 frames for all training segments (instead of random durations between 200 and 400 frames). 
    \item We sample the training segments in such a way that all regions of a recording are used equally (instead selecting the segments completely at random).
    \item  We generated around 700 Kaldi archives such that each of them contained exactly 15 training samples from each speaker (i.e. around 107K samples in each archive).
    \item The network has nine TDNN layers, which sees a total context of 13 frames per side, before the statistics pooling layer. 
    \item The input to the statistics pooling layer is the concatenation of the output from the 7th and the 9th TDNN layer.
\end{itemize}
 For more details, see \cite{landini2019but,LandiniICASSP20,DiezICASSP20}.  
 
\subsection{Baseline PLDA}
\label{sec:PLDAinit}
The baseline PLDA model is trained on x-vectors extracted from $3s$ speech segments from VoxCeleb 1 and 2 and utterance IDs combined with speaker IDs serve as the class labels. Before the PLDA training, the x-vectors are centered (i.e. mean normalized), whitened (i.e. normalized to have identity covariance matrix) and length-normalized~\cite{GarciaRomero2011lnorm}. The centering and whitening transformation are estimated on the joint set of DIHARD development and evaluation data.

\subsection{PLDA and embedding extractor initialization}
\label{sec:initialization}
The details of the construction and initialization of our probabilistic embedding extractor and the diagonalized PLDA model are given here. Recall that the embedding extractor does: 
$$s_t\mapsto\hat\xvec_t, \Bmat_t$$ 
where $s_t$ is a speech segment, $\hat\xvec_t\in\R^D$ and $\Bmat_t=\{b_{jt}\}$ is a diagonal precision matrix. The diagonalized PLDA model has a single parameter, the diagonal precision matrix, $\Wmat=\{w_j\}$. 

In this work, we build our probabilistic embedding extractor by modifying an existing, baseline x-vector extractor. The existing extractor \emph{includes} the final centering, whitening and length normalization as detailed in section~\ref{sec:PLDAinit} above. In all our experiments the x-vector extractor, including the centering, whitening and  length norm, remain fixed and are not subject to retraining. 

The modifications are as follows. First, we add a linear transformation after the standard (length-normalized) x-vector, so that transformed x-vector is our desired $\hat\xvec_t$. Second, the output of the statistics pooling layer in the x-vector extractor is passed through an additional feed-forward neural network which outputs the diagonal precisions, $\Bmat_t$. The parameters of these extra components, the linear transform and the non-linear precision extractor are subject to training in our experiments. The third trainable parameter set is the diagonal PLDA within-class covariance, $\Wmat$.

We use a diagonalizing transformation of the original baseline PLDA to initialize the linear x-vector transformation and $\Wmat$. Here, we decrease the dimensionality of the original x-vectors from 512 to 500, where we keep the 500 dimensions with the highest speaker variability.

As an extractor for the precisions, $\Bmat_t$,  we use a feed-forward neural network with a single hidden layer and the \emph{softplus} activation function. The whole net has the stucture: linear-softplus-linear-softplus. The final softplus is needed to give non-negative precisions. As an input, this network uses the output of the statistics pooling layer of the original x-vector extractor. The first-order statistics are used as they are, while second-order statistics are inverted. That is done to speed up the training since the original statistics contain standard deviation and we are interested in the precision at the output. Also, we concatenate the statistics vector with the duration of the speech segment. The resulting dimensionality of the input is 2049. The hidden layer is 1000 dimensional and the output has a size of $D=500$, to agree with the size of $\hat\xvec_t$. 

The parameters of the precision extractor network are initialized randomly in such a way that at the beginning of the training the $b_{jt} \gg w_j$ condition (see Section \ref{sec:cpost}) would be satisfied.

So, after the initialization, the model performs very similarly to the baseline PLDA with standard x-vectors.
\subsection{Parameter training}
Once the model is initialized, we train it using the cross-entropy objective from Section \ref{sec:training}. Our training examples are octets of x-vectors selected randomly from the training data. All segments in an octet come from the same recording, but it is not guaranteed that they are consecutive. Selecting the segments in a random order results in more difficult training examples and leads to faster training. (We did also try using consecutive segments and it performs similarly, except that training then takes longer.) We use \emph{stochastic gradient descent} for the training, the parameters are updated after seeing a mini-batch of 100 examples. Since we initialize the PLDA and the transformation matrix from the existing PLDA we do not want them to deviate from the initial values too quickly so that the precision extractor would benefit from this smart initialization. In order to achieve that, the learning rate for these parameters is set to a much smaller value ($10^4$ times smaller in our case) than the one for precision extracting network. 

In our experiments, for the prior $P(\Lmat)$, we used the \emph{Chinese restaurant process} (CRP)\ \cite{Pitman,fox2011sticky,zhang2019fully}. We assigned the parameters of the CRP (the concentration and discount) to control the distribution of the number of speakers: The variance is maximized, subject to the constraint that expected number of speakers for $N$ segments is the same as the true number of speakers in the training set, where $N$ is the total size of the training set. As mentioned above, the CRP gives an exchangeable distribution that is invariant to the order of the speaker labels. This prior is appropriate for our procedure of randomly selecting the segments to form the octets during training. For situations where the order matters, other priors could be considered, for example those mentioned in section~\ref{sec:prior}.

\subsection{Results}
\begin{table*}[!th]

\centering
\caption{\label{tab:results} Diarization error-rate [\%] on DIHARD 2019 development and evaluation sets. }
\begin{tabular}{l c c c c c c} 

    \toprule
	&\multicolumn{2}{c}{$\sigma$=0}	&	\multicolumn{2}{c}{$\sigma$ optimal}	&	\multicolumn{2}{c}{likelihood scaling optimal}\\	
	&dev	&eval&	dev&	eval&	dev&	eval\\
\midrule
Baseline AHC&	-&	-&	25.12&	26.33&	-&	-\\
by-the-book	AHC&44.28&	41.09&	23.82&	23.98&	22.85&	25.18\\
by-the-book AHC, PLDA trained& 26.95&	26.34 & 22.38 & 22.68 & 21.05 & 21.86\\
by-the-book AHC, PLDA + embedding extractor trained	&22.95&	23.38&	21.61&	21.84&	20.95&	21.79\\
    \bottomrule  
\end{tabular}
\end{table*}

The results are summarized in table~\ref{tab:results} in terms of diarization error-rate (DER). For each clustering strategy (rows), the table presents three sets of results (pairs of columns). The result sets differ in how the AHC stopping threshold, $\sigma$, was set. In the first two columns, $\sigma=0$, which is the optimal maximum-likelihood threshold. 

In the second set, $\sigma$ was tuned on the DIHARD development set to minimize diarization error-rate. There is an important difference between the baseline and the other systems: The scores for the merging decisions in the baseline AHC are individually calibrated for each recording, using the unsupervised calibration as explained above. The globally optimized threshold, $\sigma$, was applied to these calibrated scores. While for the rest of the experiments no such calibration was done and $\sigma$ was applied directly to~\eqref{eq:Delta}. 

Finally, to compensate for the fact that the segments in the evaluation data are not independent, one can scale down $\log P(s_t\mid \yvec)$, the speaker identity log-likelihoods of the individual speech segments~\cite{stafylakis2013compensation,kenny2010diarization}. For us that means scaling the statistics $\bar a_j$ and $\bar b_j$ in (\ref{eq:bars}). The last two columns of the table show the results with the tuned likelihood scalar. The AHC stopping threshold, in this case, is set to $\sigma=0$. One can notice that optimally scaling the likelihoods has a similar effect to optimizing $\sigma$. 

The first row in the table presents the baseline approach. The baseline is Kaldi-style AHC (section~\ref{sec:baseAHC}), with the same PLDA and x-vector extractor we use to initialize our model. Here, an important note is that principal component analysis (PCA) is applied to the x-vectors, \emph{independently} for each recording, giving a different PCA projection for each recording. The PCA output dimension is adaptively chosen so that the resulting low-dimensional projected x-vector retain only 30\% of their variability. The projected x-vectors are then length normalized again prior to the clustering. The PLDA parameters are also projected to the corresponding low-dimensional space. The resulting PLDA model is used to calculate log-likelihood ratio verification scores as the similarity metric that is applied to all pairs of x-vectors in the recording. The resulting similarity matrix is the input to the AHC. If instead, omit these extra PCA and length normalization steps (as in the rest of the experiments), the baseline results degrade to 28.63\% and 28.36\% DER on development and evaluation sets respectively.

The rest of table \ref{tab:results} shows the results for the diarization with our version of AHC (section~\ref{sec:bookAHC}), where we use by-the-book scoring at each stage of the clustering. First (table row 2), we applied AHC with exact scoring to the untrained model initialized from the baseline. As mentioned earlier, the model before training is practically the same as the baseline. So, the main source of performance difference is the alternative way of clustering. This model is heavily miscalibrated and using $\sigma=0$ as a threshold results in a performance inferior to the baseline. But, by tuning either the threshold or the likelihood scaling, one can achieve an improvement over the standard AHC clustering. 

Next (table row 3), we present the results obtained with a system where the part of the network that extracts precisions remains fixed, so that it keeps extracting large precisions, $b_{jt}$. We retrain only the PLDA (the $w_j$) and the linear transformation (in the extractor) that outputs $\hat\xvec_t$. One can see that the calibration problem in this case is not that prominent anymore. But still, there is a possibility to improve the performance by carefully optimizing the threshold or the likelihood scalar.

Finally (table row 4), we enable training of all three parameter sets: the linear transform, the precision extractor and the PLDA. Here, even without tuning of the hyper-parameters, we improve over the baseline, although threshold or likelihood scalar tuning can still give small improvements. 

An interesting observation is that as we move from untrained model to the one with PLDA training to the one with all components trained, the optimal values for the threshold are approaching zero. Similarly for the likelihood scaling factor, for each next system it is closer to $1$.  

\section{Conclusions}
We have made the following contributions:
\begin{itemize}
    \item We have presented a novel methodology, developed around the factorization~\eqref{eq:gsump}, to derive discriminatively trainable embedding extractors from generative models. 
    \item We have motivated and experimentally demonstrated arguments for data representations in the form of probabilistically distributed hidden embeddings. In contrast to variational autoencoders~\cite{VAE} and other works on probabilistic embeddings, e.g.~\cite{hib}, that extract hidden variable \emph{posterior} distributions, we extract \emph{likelihood} distributions.
    \item We have applied the above to the x-vector--PLDA cascade, showing how to adapt PLDA to score probabilistic embeddings in a computationally efficient way.  
    \item For speaker diarization, we have introduced a new discriminative training objective, a large multiclass cross-entropy, which is a proper scoring rule~\cite{PSR} that encourages well-calibrated likelihood distributions~\cite{NikoPhD} over all the $B_n$ possible ways to cluster an $n$-tuple of speech segments according to speaker. This was experimentally demonstrated on joint discriminative training of the embedding extractor and the PLDA backend.
    \item A new AHC algorithm that makes full use of the PLDA model to greedily optimize the maximum likelihood clustering hypothesis. 
\end{itemize}
In future, we would like to make the embedding extractor more powerful by making the precision extractor deeper and also retraining the parameters of the whole x-vector extractor. The extra capacity can be regularized by more aggressive augmentation of the input segments to provide a greater variety of durations, SNR and other quality factors. We would also like to try the uncertainty propagation into the BHMM backend.

\section{Acknowledgements}
The work was supported by Czech National Science Foundation (GACR) project ``NEUREM3 No. 19-26934X, European Union’s Horizon 2020 grant no. 833635 ``ROXANNE and by Czech Ministry of Education, Youth and Sports from the National Programme of Sustainability (NPU II) project ``IT4Innovations excellence in science - LQ1602.

\bibliographystyle{IEEEbib}
\bibliography{embeddings}

\appendix
\section{Discriminative disclaimer}
\label{sec:appendix}
Although our model~\eqref{dg:model1} for embedding extraction is generative, where the embeddings are supposed to have generated the observed speech, we must emphasize that our recipe is discriminative in the sense that the observed data is not fully modelled. Recall~\eqref{eq:gsump}, reproduced here for convenience:
\begin{align}
\label{eq:gsumpA}
P(s_t\mid \xvec_t) &=h(s_t)\;\times f(s_t,\xvec_t)
\end{align}  
In the discriminative training of $f$, $h$ plays no role and it fails to provide a mechanism to ensure that once $f$ has been trained, then~\eqref{eq:gsumpA} will give normalizable distributions for the observed data. Discriminative training does not guarantee that some function $h(s_t)>0$ exists that would normalize~\eqref{eq:gsumpA} for every $\xvec_t$:
\begin{align}
\label{eq:badnorm}
\int h(s_t)f(s_t,\xvec_t) \,ds_t = 1, \forall \xvec_t\in\R^D
\end{align}  
Arguably the best practical way to (approximately) enforce this constraint, is to do generative rather than discriminative training. This would typically require some \emph{variational Bayes} (VB) recipe to deal with the intractable marginalizations over the hidden variables. In VB, the hidden variable posterior is approximate and in general will also \emph{not} be exactly consistent with a normalizable generative model for the observed data. However, in this case, the training objective (the evidence lower bound, or ELBO) will at least tend to minimize this inconsistency, by effectively minimizing KL divergence from approximate to true posterior~\cite{PRML}.

Apart from stating that we \emph{do} plan to try generative VB training in future, we could motivate that the failure of~\eqref{eq:badnorm} should not trouble us any more than it seems to trouble the designers and users of any other discriminative training recipes---or indeed any recipe that extracts features without a full generative model of the observed data. Discriminative recipes generally do not define generative models for the data and even in cases where they do, discriminatively trained generative models are usually defective as models of the observed data. If we do not need $h(s_t)$ during training or runtime, should it unduly trouble us that it may not exist?

\end{document}